\documentclass[12pt]{article}
\usepackage{graphicx}
\usepackage{amssymb,amsmath,amsfonts,palatino,amsthm}
\usepackage{amssymb}
\usepackage{epstopdf}
\newcommand{\beq}{\begin{equation}}
\newcommand{\eeq}{\end{equation}}

\newcommand{\f}{\begin{equation}}
\newcommand{\ff}{\end{equation}}

\begin{document}

\title{The black hole information paradox and relative locality \\}
\author{Lee Smolin\thanks{lsmolin@perimeterinstitute.ca} 
\\
\\
Perimeter Institute for Theoretical Physics,\\
31 Caroline Street North, Waterloo, Ontario N2J 2Y5, Canada}
\date{\today}
\maketitle

\begin{abstract}

We argue that the recently proposed {\it principle of relative locality} offers a new way to resolve the black hole information puzzle.  

\end{abstract}

\tableofcontents

\newpage

\section{Introduction}

The black hole information paradox\footnote{For two recent critical review,  see \cite{sabinelee, samir}.} has challenged theorists of quantum gravity since first proposed by 
Hawking\cite{hawking}.  One much discussed view has been that some kind of non-locality is required to resolve the puzzle\cite{BHC}.  A recently proposed framework for quantum gravity phenomenology, called relative locality\cite{PRL,GRB,SBP,GRG} does feature a very controlled form of non-locality.  We argue here that the kind and scale of non-locality implied by relative locality is sufficient to resolve the black hole information paradox.  

Whatever the quantum theory of gravity that describes nature is,  we have good reason to suspect that it involves a dissolving of the usual notion of
locality in spacetime.  It is therefore of interest to characterize exactly how non-locality first appears in physical phenomena in experimental regimes where
one of the Planck scales becomes evident.  The Principle of the relativity of locality (PRL) was introduced as a possible characterization of how non-locality emerges which is relevant for a regime of quantum gravity phenomena  in which 
we can neglect $\hbar$ and $G_{Newton}$,  but where ratios of energies to the Planck mass, $m_p = \sqrt{\frac{\hbar}{G_{Newton}}}$ may still be detected.

The principle of relative locality states that in this regime there is no invariant description of events in spacetime.  The invariant description in this regime is that the motions and interactions of particles take place in a phase space, $\Gamma$,
which is the cotangent bundle over a curved momentum space\footnote{by which we mean the space of four-momenta.  Note that we use units in which 
$c=1$, but we leave $\hbar$ and $G_{Newton}$ explicit.}, $\cal P$
\f
\Gamma = T_* ({\cal P})
\ff
$m_p$ is taken as the scale of the curvature of momentum space.  
There are several consequences of the curvature of momentum space\footnote{For details,  please see the original papers\cite{PRL,GRB,SBP,GRG}.} :

\begin{itemize}

\item{}Einstein's procedure for assigning spacetime coordinates to distant events by exchanges of light signals\cite{Einstein} becomes dependent on the energy of the light signals and particles involved in the events.  Equivalently, translations from the observations made by one observer to those of a distant observer are now energy dependent.  This is due to the fact that translations are generated by conservation laws,  which are non-linear in energy and momentum due to the curvature of momentum space.  

\item{}Consequently spacetime coordinates of distant events become energy dependent in a precise way.  It can be said that, when $m_p$ is relevant, spacetime and momentum space are merged into a single invariant structure-the phase space $\Gamma$,  in the same sense in which space and time are merged into spacetime when $c$ is relevent.  

\item{}As a consequence,  the principle of locality is modified for observations of distant events.  It remains true that events local to an observer, appear to occur locally in that observer's spacetime coordinates.  However, events a distance $x$ from an observer,  involving a particle with energy $E$,  will appear to involve interactions of particles separated by\footnote{For the precise formulas see \cite{PRL,GRB}.}
distances of order of
\f
\delta x \approx |x| \frac{E}{m_p c^2}
\label{deltax1}
\ff

\item{}Mathematically, this can be understood as follows: particles with different four momenta live in different spacetimes which are sections in the bundle $ T_* ({\cal P})$ over the corresponding points in $\cal P$.  Particles moving on different sections of $T_* ({\cal P})$ associated with different momenta
interact under a notion of locality according to which the ends of the world lines of particles with different momenta meet each other under parallel transport on
the curved momentum space.  This is as local as interactions can be in this context.

\end{itemize}

The same apparent non-localities affect the description by an observer of events far to the future or far to their past, by a time $t$ measured by their clocks. 
\f
\delta x \approx |t| \frac{E}{m_p c^2}
\label{deltat1}
\ff

\begin{figure}[h!]
\begin{center}
\includegraphics[width=0.5 \textwidth]{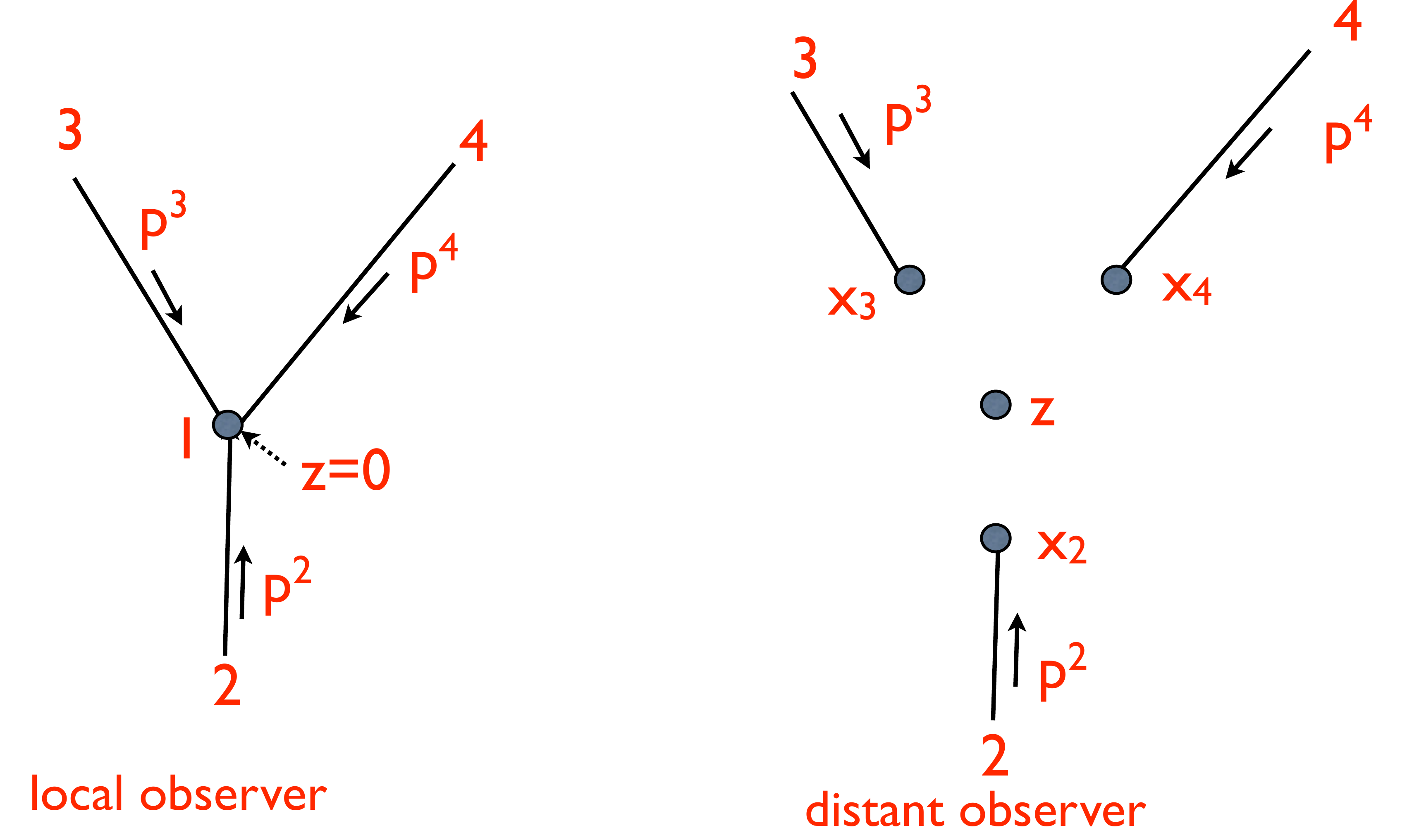}
\end{center}
\label{relloc1}
\caption{In relative locality,  a  local interaction appears local to observers local to it, but appears spread out using the coordinates of distant observers.}
\end{figure}

The apparent non-locality due to relative locality are observer dependent, as illustrated in Figure (\ref{relloc1}).  For single events such as pictured, the non-localities can be transformed away by translating to the frame of an observer local to the event.  There are real physical non-localities, but they only show up in processes involving widely separated events so that no local observer can transform away all the non-locality.   Consequently, in spite of the fact that observers distant from each other can disagree about which events appear local and which appear non-local, there appears to be no physical paradoxes or contradictions of the kind suggested in \cite{unruh-paradox,sabine-paradox}.

To discuss the details one has to consider the various ways in which the momentum space may be curved.  We will assume here that, as discussed in detail in \cite{PRL,GRB}, there is non-metricity and/or torsion at order $\frac{1}{m_p}$, which yields effects of order (\ref{deltax1}, \ref{deltat1}).  This is the case in the best studied example, which is physics in the non-commutative $\kappa$-Minkowski spacetime\cite{kP-RL,kP-all,limitations}.  

An example of how relative locality can lead to real physical effects occurs in the analysis of anomalous time delays in gamma-ray bursts (GRB)s, as was discussed in detail in\cite{GRB,kP-all}.  Two photons of different energies are observed to be emitted by the source.  They are observed by an observer local to the emission events to be emitted simultaneously.  They travel for on the order of $10$ billion years till they are detected by a detector on the Fermi satellite.  The detection events are local when described by an observer on the satellite.  

In the analysis of this process a gauge can be chosen so that the speed of light is energy independent.  (This fixes a gauge freedom that arises by virtue of the freedom to choose coordinates on the curved momentum space.)  Nonetheless all observers agree there is a time delay between the times they are registered by detectors on the satellite.  According to the observer at the emission, this is due to non-locality at the detectors.  According to an observer at the detector, the non-locality occurred in the emission events.  Remarkably they all agree on the amount of time delay to be observed.

As we will now show, a similar analysis applies to the process of black hole evaporation.  The key point is that the evaporation takes a sufficiently long time that no observer sees the physics of the formation of the black hole and final evaporation as both local.  Due to (\ref{deltat1}) one event or the other, or both, will be described in the coordinates of any local observer, to be sufficiently non-local as to evade the paradox.  

In the next section we first give the usual argument for the black hole information loss paradox and then show how it is resolved from the viewpoint of two observers, one near the black hole at the time of its formation, the second nearby at the time of final evaporation.  




Before giving the argument, I should caution that it works at a very heuristic level of (non)-rigor.  In particular, relative locality has yet to be formulated in detail in curved spacetime.  I then use the formula (\ref{deltax1},\ref{deltat1}) in curved spacetime, whereas it has only been derived in flat spacetime, in the approximation where effects of gravity, proportional to $G_{Newton}$  have been ignored.  To apply to black holes, we have to assume that processes that appear to be non-local and to involve interaction of particles that appear to be outside of each other's lightcones in flat spacetime, will behave in the same non-local fashion,  even when the non-local interaction crosses the 
apparent horizon of an evaporating black hole.  This would seem difficult to avoid,  as the geometry near the horizon is approximately that of flat Rindler spacetime.

\section{A possible resolution of the BH information paradox}

\subsection{The usual statement of the black hole information paradox}

We first state the black hole information problem. Consider a black hole of mass $M >> m_p$ which forms at time $t_0$ in an asymptotically flat\footnote{It will not change the argument to assume the spacetime is asymptotically AdS.}   spacetime.   It can contain up to
\f
N=M^2/m_p^2
\ff
bits of information. One can see that by forming the black hole by dropping in the $N$ bits of information one at a time\cite{bekenstein}.  In this process a box with some bits of information is slowly lowered down to the surface of the black hole, then emptied when it touches the horizon. The box itself can be macroscopic, all that is required is that the size of the box, $R_{box} < R_{Schw} = 2GM$.

We will consider the case where $N$ bits of information have been added in this way.

The black hole evaporates for a time
\f
t_e = t_p \frac{M^3}{m_p^3}
\ff
till there remains a quantum black hole, which is an object with unknown properties, with a mass of $m_e \approx m_p$ and a radius $r_e \approx l_p$.  There also is present a bath of thermal radiation with entropy $S_r \approx N$, spread out in a sphere of radius $t_e$.  

Let us assume that there is not a permanent remnant and also that the problem is not resolved by the formation of a baby universe\footnote{We ignore these possible resolutions as we are interested in a possible resolution where the black hole evaporates completely.}. 
If so then the black hole has to eventually disappear by a time $t_f > t_e$.  
We also assume that the problem is not resolved by some unknown process which leaks the information out during the semiclassical evaporation process.  This means that at $t_e$ the quantum black hole still contains almost all of the $N$ bits of information.  The reason is that we can assume that the black hole formed initially in some pure state, and it is thus now entangled with the thermal radiation in an overall pure state.  This implies that the information contained in the black hole is enough, when combined with  the thermal radiation, to reconstitute a pure state.  This means it must contain an amount of information equal to the entropy of the radiation, which is $N$.  

But if unitarity in the asymptotically flat spacetime is to be maintained,  then there must be some process by which the $N$ bits of information are to be released or leaked by the quantum black hole between the times $t_e$ and $t_f$.   This requires the production of $N$ quanta, each with a tiny energy
\f
\epsilon_r = \frac{m_p}{N}
\ff
which means each has a wavelength
\f
\lambda_r = N l_p = l_p \frac{M^2}{m_p^2}
\ff
How could an object with a radius of the Planck length couple to such enormous wavelengths?  It would seem to require a large amount of non-locality to accomplish this.  

We can put the point more operationally as follows.  An observer at the time of formation of the black hole can create a quanta just outside the horizon, with a momentum towards the horizon that will, assuming locality, inject it into the horizon. $N$ bits of information can be coded and injected this way.   If the evolution is unitarity and the black hole evaporates completely, an  observer at a time later than $t_f$ has to be able to detect that quanta.  This seems unlikely if physics is local because the quanta is at time $t_e$  trapped into a region of Planck size that cannot couple effectively to long wavelength modes.  But the final quanta must have a wavelength $\lambda_r$ if all the $N$ bits of information are all to be detectible, which is necessary if the evolution is unitary.  

\subsection{The point of view of an observer at the formation of the black hole}

Relative locality changes this argument crucially. Let us consider first the point of view of an observer present at the formation of the black hole. Such an observer describes accurately the process by which the boxes containing the $N$ bits of information are lowered into the black hole.  As they are near to that process they have no problem identifying when the boxes meet the black hole horizon and drop their information into the horizon.  

Let us follow our observer as he predicts the future evolution of  one bit of that information, encapsulated in a quanta with energy $E$ (as measured by that observer) that falls through the horizon at time $t_0$.  Let us choose that bit to be one injected late in the formation process, when the black hole already had mass $M$.   
As he must stay outside the horizon, he sees it at any time  $t > t_0 $ to be a component of the quantum state of the  black hole.  However, by that time, the uncertainty in his ability to locate that quanta is given by (\ref{deltat1}),
\f
\delta x \approx  t \frac{E}{m_p}
\ff
This means that a detector set up to detect a quanta of that kind would have an equal probability of detecting it anywhere within a radius of $\delta x$ of the detector.
However, he knows that the quanta is initially confined to within the horizon of the black hole, so by the Heisenberg uncertainty principle, 
\f
E > \frac{\hbar}{GM}
\ff 

Thus, at a time $t$ the relative locality uncertainty in the position of the quanta is
\f
\delta x >   t \frac{m_p}{M}
\ff
This grows in time untill a time $t_{rl}$ when the relative locality uncertainty exceeds the Schwarzchild radius, $
\delta x > 2GM$. This is at a time 
\f
t_{rl} = t_p \frac{M^2}{m_p^2}
\ff
This is a long time, but much before the evaporation time $t_e$.  Between $t_{rl}$ and $t_e$ the quanta in the black hole are seen by the formation-time observer to interact with particles or detectors outside the horizon. The distance $\delta x $ over which a detector external to the horizon will still interact 
with the quanta grows until, at the evaporation time, it is 
\f
\delta x_e >   t_e  \frac{m_p}{M} = l_p \frac{M^2}{m_p^2}
\ff

Thus from the point of view of the observer at formation the quantum black hole at time $t_e$ behaves as if it were  a very non-local object.  If we describe it operationally, the formation time observer would predict that a detector placed within a radius of $\delta x_e$ at time $t_e$ or later is able to detect the quanta.  Thus, from
an operational point of view, the quanta making up the quantum black hole cannot be  considered to be confined to
a region of Planck size.  Instead, the quanta making up the black hole respond to detectors as if they were spread over a region of size $\delta x_e$.  Notice that $\delta x_e = \lambda_r$.  
 There is then no problem 
coding each bit of the original information to a quanta with wavelength $\lambda_r$ because the observer at formation sees each bit of the original information to already 
occupy a region of that size.  So relative locality provides exactly enough non-locality to resolve the black hole information problem from the point of view
of the observer at the black hole's formation.

\subsection{The point of view of the observer at the final evaporation}

Let us now consider the point of view of an observer at the time of evaporation, $t_e$ who is within a distant $\lambda_r$ of it.  Because locality is relative, and observers see physics nearby them to be local, the evaporation-time observer sees a quantum black hole with about the Planck mass taking up a region of radius about the Planck length.  Because she sees physics in her vicinity to be  local, she knows she cannot easily detect any quanta trapped inside the Planck size quantum black  hole a distance
$\lambda_r$ from her.  Additionally, for her there is a problem
of how could $N>> 1 $ bits of information be packed into an object so small.   

To study the problem she follows one of the $N$ bits of information back to the formation of the black hole at time $t_0$.  She knows there was a process by which that quanta of information was coded into a photon initially trapped in a box which was then brought near to the black hole to be dropped in through the horizon.  She knows this because she will have a record of this sent to her by the formation time observer. 
So she attempts to reconstruct a description of the process by which that bit, and the other $N-1$ bits of information were dropped into the black hole horizon at time $t_0$.

However, her attribution of spatial position to the quanta in question at time $t_0$ are limited by the same relative locality uncertainty
\f
\delta x_i  =  l_p \frac{M^2}{m_p^2} >> R_{Schw} = 2 GM
\ff
because we assume the initial black hole is large in Planck units.  Hence in her description, the quanta that was released by the box interacts as if it is a non-local object, spread out over a region much larger than the Schwarzchild radius of the black hole.  Hence it is not probable that the quanta ended up within the black hole's horizon.  Instead, when the box releases the quanta it will be seen by her to be anywhere within a radius of $\delta x_i = \lambda_r$.  

Hence the evaporation-time observer must say that the quanta was likely never contained within the black hole horizon, even at the beginning.  From her point of view, the black hole composed of the $N$ quanta released at the horizon is from the very beginning a non-local object, spread out over a radius of $\lambda_r =  l_p \frac{M^2}{m_p^2} $.  It is no surprise to her that a long time later the quanta that were released at the horizon may be detected within the same large radius.  

Note that the issue is not whether the box was lowered to the horizon at time $t_0$.  This is because the box itself can  be considered to be  a macroscopic object.  
As was shown in \cite{SBP}, macroscopic bodies made out of $N_c$ elementary particles live on a curved momentum space whose radius of curvature is
$N_c$ times that of the elementary particles.  Hence large composite bodies behave to an excellent approximation as if their momentum spaces are flat and they hence are well described by special relativity.   So all observers will agree that the box was lowered to the black hole horizon before being opened. 
From the perspective of a formation time observer, the quanta is in the box and is ejected through the horizon.  But from the point of view of an observer at a much
later time, $t_e$ the quanta cannot be considered to be confined to the box because its relative locality uncertainty is $\delta x_i =  l_p \frac{M^2}{m_p^2} >> 2GM > R_{box}$.  

The problem with observers disagreeing about whether a microscopic quanta is contained within a macroscopic box was pointed out by Schutzhold and 
Unruh\cite{unruh-paradox} some time ago in the context of deformed (doubly) special relativity.
Relative locality tells us that there is no paradox, so long as one reasons consistently to the different descriptions of different observers.  

\section{Conclusion}

In previous studies of relative locality\cite{PRL,GRB,SBP,GRG} it was seen how two observers, distant from each other, can end up explaining the same objective phenomena with two pictures that are seemingly rather different.  This is certainly the case here.  

What all observers agree about is the initial and final observations.  In the case of the black hole all observers agree that a quanta which was initially contained in a box of size $L < GM$ is released when the box is held up against the horizon of a large black hole, with mass $M > > m_p$.  All observers also agree on what happened an evaporation time, $t_e$ later:  that or a corresponding quanta was detected with an energy $\epsilon = m_p \frac{m_p^2}{M^2}$, and with an equal probability to be detected by a detector anywhere within a volume surrounding the residue of the black hole of radius $\lambda_r = l_p  \frac{M^2}{m_p^2}$.   

 Note that the horizon is not a physical barrier.   If relative locality means that very distant  interactions appear to take place outside the lightcone of the particles involved in the interaction in flat spacetime, this must be the case in a black hole spacetime as well.  So from the point of view of the formation-time observer the potential paradox is resolved because relative locality forces them to the conclusion that a quanta initially inserted inside a black hole horizon can be detected by a detector outside the horizon a time $t_{rl} =  t_p \frac{M^2}{m_p^2}$ later, which is long before the evaporation time.  Thus, for the bulk of the evaporation time,  $t_{e} =  t_p \frac{M^3}{m_p^3}$. the formation-time observer sees quanta initially inserted inside the horizon able to interact with matter outside the horizon. 

The evaporation-time observer tells the story a different way.  The quanta in the box can, when interacting with the box, appear to jump a distance $\lambda_r$.  Note that the interaction with the box is necessary for the box to expel the quanta to begin with.  Thus, the quanta almost certainly starts off outside the black hole horizon.  
After a time $t_e$ it is detected by a local observation to be still outside the horizon.  Thus the evaporation-time observer has no paradox to answer because in their records the quanta was never inside the horizon.  

In fact for most of the evaporation time the two observers agree that the quanta could be detected by detectors outside the horizon.  

We close with some further comments.

\begin{itemize}

\item{} Relative locality is a framework to explore the consequences of curved momentum space. It describes the consequences of torsion, curvature or non-metricity in momentum space.  It does not predict whether they are present, that could be accomplished by deriving relative locality as an approximation to a full fledged quantum theory of gravity.  Here we have assumed the largest leading order effect which is that the torsion and/or non-metricity of momentum space are of order $\frac{1}{m_p}$.  If these effects are not present the next leading order involves curvatures which are naturally of order $\frac{1}{m_p^2}$.  In this case the estimates for 
$\delta x_e$ shrink to $GM$.  This is still enough to resolve the information paradox, but not so comfortably.  

\item{} It must be emphasized again that this is an heuristic argument, pending the extension of relative locality to general relativity and curved spacetime.  Usual general relativity describes a phase space as ${\cal T}_* ({\cal M})$, the cotangent bundle of a curved phase space manifold $\cal M$.  Relative locality describes physics in an opposite regime, where the phase space is the cotangent bundle of a curved momentum space, $\Gamma = {\cal T}_* ({\cal P})$. In this case spacetimes (labeled by momenta) are cotangent planes and hence are flat.  We require the description of physics when both spacetime and momentum space are curved.  This is equivalent to turning on $G_{Newton}$ and $\hbar$.  

\item{} The reasoning here can be compared to the proposal of  black  hole complementarity\cite{BHC}.  There are differences, in particular, here all observers agree about the initial and final events, but just disagree on the process between them.  
Nonetheless, the  possible relationship between them needs to be explored.

\end{itemize}

\section*{ACKOWLEDGEMENT}

Many thanks to Laurent Freidel for a conversation at an early stage of this idea.   Thanks also to Daniel Gottesman for an encouraging remark about this idea and to  Giovanni Amelino-Camelia ,  Jerzy Kowalski-Glikman and Samir Mathur for very helpful comments on the manuscript.   Research at
Perimeter Institute for Theoretical Physics is supported in part by
the Government of Canada through NSERC and by the Province of
Ontario through MRI.

\end{document}